\documentclass[universe,article,accept,pdftex,moreauthors]{Definitions/mdpi} 
\graphicspath{{figs/}}
\usepackage[T1]{fontenc}
\usepackage{xcolor}
\usepackage{amsmath}
\usepackage{graphicx}
\newcommand{\eb}{\begin{equation}}
\newcommand{\ee}{\end{equation}}

\definecolor{rkka}{RGB}{219,66,32}

\firstpage{1} 
\pubvolume{10}
\issuenum{1}
\articlenumber{15}
\pubyear{2024}
\copyrightyear{2023}
\externaleditor{Academic Editor: Giacomo Tommei}
\datereceived{30 November 2023} 
\daterevised{23 December 2023} 
\dateaccepted{26 December 2023} 
\datepublished{28 December 2023} 
\hreflink{https://\linebreak doi.org/10.3390/universe10010015} 

\Title{Chaotic Capture of a Retrograde Moon by Venus and the Reversal of Its Spin}

\TitleCitation{Chaotic Capture of a Retrograde Moon by Venus and the Reversal of Its Spin}

\Author{{{Valeri}
} 
 V. Makarov $^{1,}$*\orcidA{} and Alexey Goldin $^{2}$}

\AuthorNames{Valeri V. Makarov and Alexey Goldin}

\AuthorCitation{{Makarov,} 
 V.V.; Goldin, A.}

\address{%
$^{1}$ \quad United States Naval Observatory, 3450 Massachusetts Ave. NW, Washington, DC {20392,} 
 USA\\
$^{2}$ \quad Teza Technology, 150 N Michigan Ave, Chicago, IL {60601}, USA; alexey.goldin@gmail.com}

\corres{Correspondence: valeri.makarov@gmail.com}

\abstract{Planets are surrounded by fractal surfaces (traditionally called Hill spheres), separating the inner zones of long-term stable orbital motion of their satellites from the outer space where the gravitational pull from the Sun takes over. Through this surface, external minor bodies in trajectories loosely co-orbital to a planet can be stochastically captured by the planet without any assistance from external perturbative forces, and can become moons chaotically orbiting the planet for extended periods of time. Using state-of-the-art orbital integrators, we simulate such capture events for Venus, resulting in long-term attachment phases by reversing the forward integration of a moon initially attached to the planet and escaping it after an extended period of time. Chaotic capture of a retrograde moon from a prograde heliocentric orbit appears to be more probable because the Hill sphere is almost four times larger in area for a retrograde orbit than for a prograde orbit. Simulated capture trajectories include cases with attachment phases up to 860,000 years for prograde moons and up to \mbox{370,000 years} for retrograde moons. Although the probability of a long-term chaotic capture from a single encounter is generally low, the high density of co-orbital bodies in the primordial protoplanetary disk makes this outcome possible, if not probable. The early Venus was surrounded by a dusty gaseous disk of its own, which, coupled with the tidal dissipation of the kinetic energy in the moon and the planet, could shrink the initial orbit and stabilize the captured body within the Hill surface. The tidal torque from the moon, for which we use the historical name Neith, gradually brakes the prograde rotation of Venus, and then reverses it, while the orbit continues to decay. Neith eventually reaches the Roche radius and disintegrates, probably depositing most of its material on Venus' surface. Our calculations show that surface density values of about 0.06 kg m$^{-2}$
for the debris disk may be sufficient to stabilize the initial chaotic orbit of Neith and to bring it down within several radii of Venus, where tidal dissipation becomes more efficient.}


\begin{document}

\section{Introduction}
Like many processes in complex dynamical systems with multiple interacting components, the formation of a planetary system is a game of probabilities. The~current structure and composition of planets, as~well as the orbital characteristics of bodies in the Solar system are the result of innumerate chances and turns stochastically taken in the garden of forking paths. At~the early stage of Solar system formation, thousands of planetesimals were roaming the protoplanetary disk in intersecting and chaotically changing orbits. Some of these primordial bodies aggregated together, merging into larger objects that would attract more of the primordial material---a gradual process leading to the formation of planets. Most of these early planets could not survive and were either ejected out of the Solar system by more massive rivals  or~recycled~\cite{2004ApJ...614..497G}. Only those planets have remained to this day that randomly found a niche in the harsh dynamical environment. The~Moon, which is apparently important for making the Earth habitable and friendly to biological life, was presumably formed in one of such dramatic catastrophes during the early ages of the system. According to the widely accepted Lunar formation theory, the~Moon accreted from a massive cloud of debris thrown up to a high orbit of several Earth's radii by a colliding body of planetary mass. The~fact that this possibly happened with the planet we live on testifies to a high density of freely roaming and mutually interacting planetesimal cores, because~the probability of a direct collision is relatively quite small compared to a fly-by~encounter.

All Solar system planets have single or multiple moons except Mercury and Venus. The~numerous moons orbiting the outer planets were presumably formed via different processes. They are believed to be either formed by self-coalescence from dusty disks surrounding the protoplanets, or~by capture from the exterior parts of the system. The~latter scenario is only viable if the phase density of such random and freely roaming minor planets is sufficiently high. Why do not the two inner planets have any satellites today? The first intuitive idea is that the proximity to the Sun may be responsible. In~a hierarchical three-body system, a~planet is in the center of an imaginary volume traditionally called the Hill sphere, within~which the gravitational pull from the planet combined with the centripetal acceleration away from the Sun is greater than the pull from the Sun. A~test particle inside the Hill sphere may remain attached (with suitable initial conditions) to the planet. The~radius of this sphere is approximately proportional to the orbital separation of the planet from the Sun. Mercury, the~closest planet to the Sun, has the smallest zone of gravitational influence, which limits its ability to retain a permanent satellite. The~second smallest Hill sphere is that of Mars, closely followed by Venus. Mars harbors two small moons, while Venus has~none.

The fate of smaller bodies approaching planets  at sufficiently small separations has been studied in numerous publications. Most of such rendezvous result in a deflection of the body's trajectory, and~statistically, in~dynamical scatter and phase mixing. But~if the angle and relative velocity of the encounter are right, there is a finite probability of two distinct outcomes: a collision with the planet and a complete ejection from the Solar system~\cite{1975AJ.....80..145W}. Each planet generates its own annular chaotic zone where smaller bodies cannot remain forever. The~radial dimension of this zone depends in a nontrivial fashion on the relative mass of the planet and the Sun~\cite{2015ApJ...799...41M,2020AstL...46..774D}. Generally, the~probability of collision is a few orders of magnitude smaller than the probability of ejection. The~formation of the Moon must then be the outcome of a rather unlikely event, with~thousands of smaller bodies ejected out of the system, and~an even greater number scattered in a wider range of orbits. A~third scenario has recently emerged, where the approaching body enters the chaotic zone around the Hill surface and becomes captured by the planet~\cite{2003Natur.423..264A}. This capture phase is transient, because~the body, while remaining in the vicinity of the planet for a period of time, moves chaotically in loops with rapidly changing planet-satellite separation. Upon~a random close approach to the planet, the~temporary satellite is eventually ejected back into a prograde heliocentric orbit. This outcome may be prevented, however, if~the orbit of the captured body becomes shrunk and regularised by efficient dynamical friction. The~excess of orbital momentum can be removed from the newly formed planet-moon system by, for~example, the~dynamical friction in the remaining debris disk surrounding the~planet.

In this study, we set out to investigate how long a stochastically captured moon can remain in the vicinity of Venus in the absence of external damping forces. This is done separately for two possible capture scenarios. The~moon can be captured into a prograde (aligned with the orbital momentum of Venus) or retrograde orbit. The~capture phase duration is then compared with the characteristic times of dissipative orbital decay due to tides in Venus. To~find a specific set of initial parameters for a successful and long-lasting capture from an external heliocentric orbit, we employ the technique of time reversal in the numerical integration of disruption events~\cite{2012MNRAS.421L..11M} when a moon initially situated in the proximity of the Hill surface becomes detached from the planet and begins to travel on its own trajectory around the Sun. We further investigate the dynamical friction of a captured retrograde moon in the planetary primordial debris disk and estimate its surface density required to efficiently reduce the planet-moon separation on the time scale of the chaotic attachment~phase.

\section{Reversal of Disruption~Trajectories}

\begin{figure}[H]
\includegraphics[width=180pt]{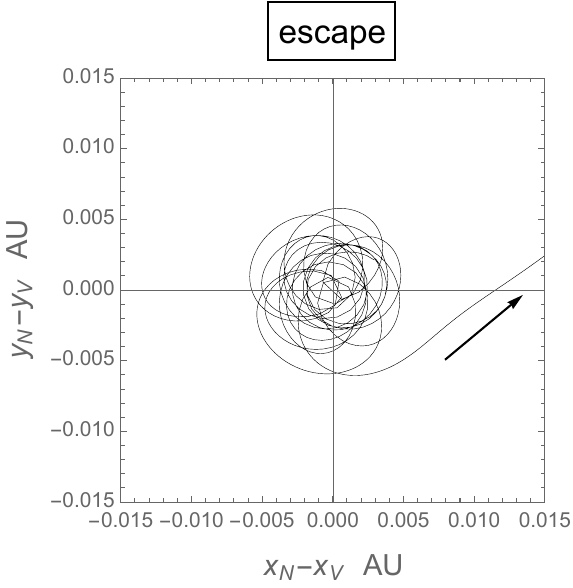}
\includegraphics[width=180pt]{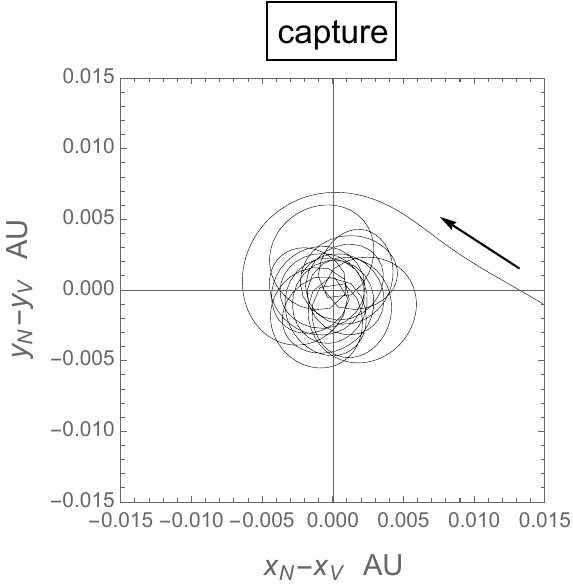}
\caption{Numerically simulated trajectories of chaotic escape (the left-hand plot) and capture (the right-hand plot) of a prograde moon through the Hill surface around Venus. The~arrows show the direction of the moon's motion with respect to Venus, which is at the center of each plot, when the moon is gravitationally detached from the~planet.} 
\label{pro.fig}
\end{figure}

It is relatively easy to find a set of initial conditions in a pure three-body problem \mbox{(Sun plus Venus plus Neith)}, resulting in a disruption of the Venus--Neith system. It is sufficient to put the moon with a finite initial eccentricity close to the Hill radius. The~values of Hill radii for prograde and retrograde eccentric orbits have been estimated via numerical simulations in the planar case~\cite{2006MNRAS.373.1227D}, when the initial orbital inclination is set to zero. Our analysis, however, is more general and three-dimensional, and~the initial orbit of Neith is allowed to be inclined to the orbit of Venus around the Sun. We find that with nonzero initial inclinations, the~effective Hill radii are somewhat smaller than the published two-dimensional estimates. Of~special interest for this study are long-lasting attachment phases, when Neith remains gravitationally bound to Venus for extended periods of time. These are found in the relatively narrow transition zone separating the inner domain for permanently stable configurations from the outer space where Neith becomes detached from Venus almost immediately or after only a few loops around Venus. Figure~\ref{pro.fig}, left-hand window, shows the final segment of one such long-lasting trajectory in a projection on the $\{x,y\}$ plane of the Venus-centric coordinate frame.
The displayed part is 2.5 years long. Neith describes chaotic, rapidly changing loops going in and out of the nominal prograde Hill sphere (which has a radius of approximately 0.0033 AU, or~81 Venus radii), sometimes coming quite close to the planet. The~relative motion of Neith in this projection is counterclockwise, i.e.,~prograde, because~it is aligned with the right-hand rotation of Venus around the Sun. The~salient feature of this particular simulation is that the attachment phase is 850,000 years long. This is how long it took Neith to finally approach Venus too close and { be} ejected as shown with an arrow in the~graph.

In order to find a corresponding capture trajectory in reversal, it is not correct just to take the end-point of the escape simulation and reverse the velocity vector, because~that would correspond to an initially retrograde (clockwise in this projection) motion of the satellite. A~symmetric transformation of the input coordinates and velocity components is required. Furthermore, even a very small change of the initial values results in a drastically different result because of the chaotic nature of the simulated process and the high Lyapunov exponential (short Lyapunov time). However, it was not too difficult to find a set of initial conditions for a body approaching from the outside that becomes captured into a chaotic prograde orbit. One such simulation resulting in a capture for 848,000 years is shown in Figure~\ref{pro.fig}, right window. Only the initial segment of the trajectory covering \mbox{2.5 years} is~plotted.

The important feature emerging from these simulations is that both escape and capture events require a sufficiently close approach of the moon to Venus. The~encounter triggering the escape or capture is as close as approximately 10 Venus radii. This is still much higher than the Roche radius of Venus, which amounts to 2.56 Venus radii. If~the intruding body approaches Venus at a closer impact parameter than the Roche radius, it is likely to be crunched by the tidal deformation force. In~the captured phase, the~trajectory remains chaotic with rapidly changing Venus--Neith separation in the range between 10~and 150~Venus radii. We note that the upper boundary of this trajectory is much higher than the estimated Hill radius for prograde orbits~\cite{2006MNRAS.373.1227D}, which is 81 Venus radii, assuming a zero initial eccentricity of the~satellite.

\begin{figure}[H]
\includegraphics[width=190pt]{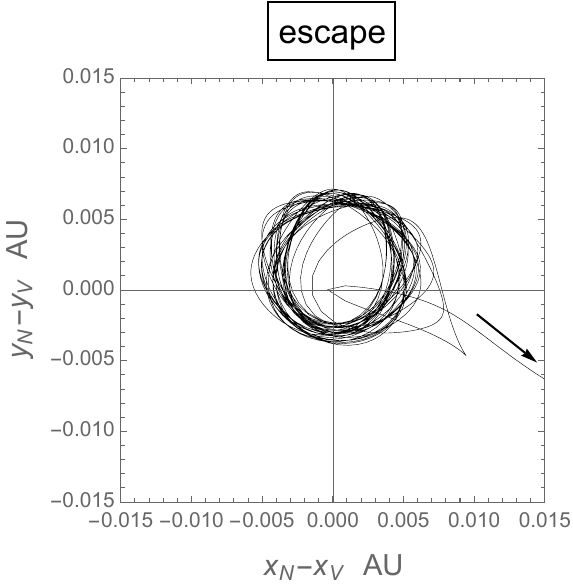}
\includegraphics[width=190pt]{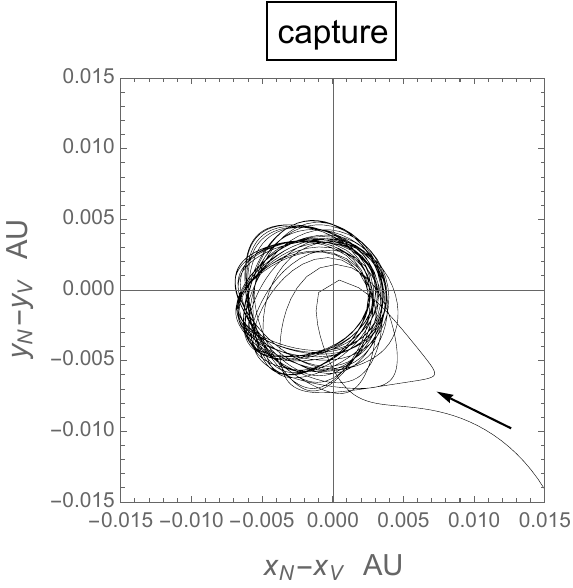}
\caption{{Numerically} 
 simulated trajectories of chaotic escape (the left-hand plot) and capture (the right-hand plot) of a retrograde moon through the Hill surface around Venus. The~arrows show the direction of the moon's motion with respect to Venus, which is at the center of each plot, when the moon is gravitationally detached from the~planet.} 
\label{ret.fig}
\end{figure}

\begin{figure}[H]
\includegraphics[width=260pt]{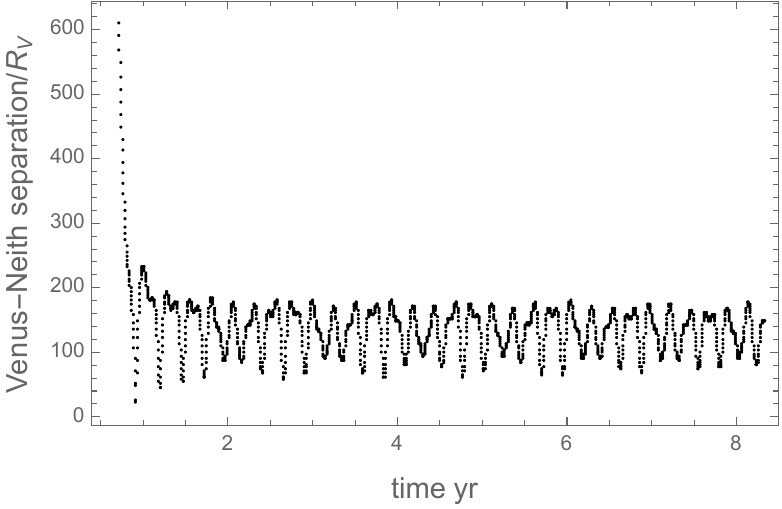}
\caption{Separation between Venus and Neith in units of Venus' radius $R_{V}$ during the first several years after chaotic capture into a retrograde~orbit.} 
\label{sep.fig}
\end{figure}

Using the same technique, we can generate three-body simulations, in~which an external planetesimal initially in a heliocentric orbit approaches Venus and becomes entrapped within its sphere of gravitational influence for a long time, while revolving around the planet in the opposite direction. We start with a satellite chaotically orbiting Venus in the vicinity of its retrograde Hill radius, which is approximately 154 Venus radii. Adjusting the initial velocity vector and the distance from the planet, we can find trajectories that are rapidly departing from Venus' vicinity, while others remain intact for extended periods of time. Figure~\ref{ret.fig}, left window, shows the end portion of one such simulated trajectory, which remained bound to Venus for 370,000 years. The~loops described by Neith are clockwise because~the orbit is retrograde with respect to Venus' orbit around the Sun. The~graph shows that the escape is triggered by a particular event, which is similar to the gravity assist maneuver used by humans to accelerate a spacecraft using the Hill sphere of a planet. The~ending portion of the trajectory sends Neith far outside the Hill radius, where it is almost stalled precipitating a nearly direct fall onto the planet. The~last close encounter at a distance of 10 Venus radii endows Neith with  momentum sufficient to leave the host planet. A~similar process in reverse is seen in the capture simulation in the right window of Figure~\ref{ret.fig}. The~initial close encounter with the planet at approximately 23~Venus radii sends Neith far above the Hill radius, where it loses some of the excess angular momentum and begins to orbit Venus in a relatively stable trajectory. The~retrograde loops in the captured phase are in fact tighter with a spread of separations between 60 and 180~Venus radii, as~shown in Figure~\ref{sep.fig}, reflecting a less chaotic~trajectory.

Our numerical simulations demonstrate that the stochastic capture of external bodies with a long-lasting bound state is possible, in principle, even without assistance from dissipative forces. But~how probable are they? Certainly, the~probability of such an outcome from a single approach is quite small, because~the chaotic capture zone covers a tiny fraction of the parameter space. However, given the high density of planetesimals in the primordial disk and the fact that each body roughly co-orbital with Venus has a large number of trials, the~cumulative probability should be significant. This is seen from the clean environment in the inner Solar system today, which lacks any significant celestial bodies apart from the four terrestrial planets. The~inner protoplanetary disk has been efficiently cleared out by the surviving planets. What happens with Neith in our simulated escape events after it departed Venus? It acquires a chaotic heliocentric orbit of its own, which results in repeated encounters with Venus. Close encounters may result in scattering to larger distances from the Sun or in direct collision with Venus. Dedicated studies of this stochastic evolution have shown that smaller-mass planets with relatively narrow Hill spheres, such as Venus, are more likely to collide with the loosely co-orbital bodies than to launch them onto higher orbits~\cite{2015ApJ...799...41M}. Thus, most of the loose co-orbital bodies must have been swept up by Venus. On~the other hand, the~cross-section of the chaotic capture zone is comparable to the cross-section of the planet if the effective width of this zone is approximately equal or greater than 1/81 (for prograde capture) or 1/154 (for retrograde capture) Venus radii. Under~these conditions, there should be as many trials of chaotic capture as direct collisions, which are believed to be responsible for the clearing out of the primordial~disk.

\section{Permanent Capture by Tidal~Dissipation}
The orbit of a chaotically captured satellite is inherently unstable in the three-body system. This phase ends with a random close approach to Venus, which provides a sufficiently high escape velocity. The~characteristic time of the capture phase is practically impossible to estimate because of the chaotic nature of the process. The~capture phase illustrated in Figure~\ref{ret.fig} lasted for 370,000 years, but~other simulations with slightly changed initial conditions resulted in much shorter times. It appears that the distribution of capture phase duration is an asymmetric heavy-tailed probability density function, and~the shown example is one of the less probable outcomes. The~temporary (captured and then detached) body  remains in the region of intersecting orbits, and~inevitably will encounter Venus multiple times, giving it more chances to be captured again, collide with the planet, or~be thrown out to greater heliocentric~distances.

Conceivably, a~dissipative agent can remove the excess orbital momentum (with respect to Venus) from a captured Neith and make it a permanent moon. Here we consider the possible role of tidal dissipation of orbital energy and compare the characteristic orbital decay timescales with the estimated capture duration. The~tidal bulge raised by Neith on Venus is not strictly aligned with the Venus--Neith vector if the mean motion of the moon is not equal to the rotational frequency of the planet. A~network of possible scenarios emerges for the long-term (secular) evolution of the orbit~\cite{2023A&A...672A..78M}. Only one of them is applicable for a retrograde Neith, because~the time lag between the tidal reaction and the variable perturbing force results in a deformation geometrically lagging the direction to the moon. This misalignment gives rise to a time-variable, but~secularly negative (retrograde) torque acting on Venus. This torque brakes the initially prograde rotation of Venus. A~symmetric couple torque is applied to Neith braking its retrograde orbital motion. Given enough time, this non-conservative dynamical interaction ends with a Venus rotating at a much slower rate in the prograde sense, or~rotating in the retrograde sense, which is observed today. Neith spirals in and may crush into Venus before the planet is fully synchronized by its~moon.

The rate of orbital decay for a non-synchronously rotating planet in the quadrupole approximation to the fourth power of eccentricity is~\cite{2019CeMDA.131...30B}:
\begin{eqnarray}
\label{da.eq}
    \frac{da}{dt} & \simeq & n\,\frac{M_N}{M_V} \frac{R_V^5}{a^4}\, [-3 (1-5 e^2+\frac{63}{8} e^4) K(2 n-2\Omega) \\ \nonumber
    && -\frac{3}{8}e^2(1-\frac{1}{4}e^2)K(n-2\Omega)-\frac{9}{4}e^2(1+\frac{9}{4}e^2)K(n) \\ \nonumber
    && -\frac{81}{8}e^4 K(2n) -\frac{441}{8}e^2(1-\frac{123}{28}e^2)K(3n-2\Omega) \\ \nonumber
    && -\frac{867}{2}e^4 K(4n-2\Omega)],
\end{eqnarray}
where $n$ is the (negative) orbital mean motion, $a$ is the semi-major axis, $R_V$ is the Venus radius, $e$ is the orbital eccentricity, $M_N$ is the mass of Neith, $M_V$ is the mass of Venus, $\Omega$ is the angular rate of Venus' rotation, and~$K(\nu)$ is the frequency-dependent quality function, also called tidal kvalitet in some papers. We are considering a prograde rotation (i.e., positive $\Omega$) for the early Venus, and {initially,}~$|\Omega|>>|n|$.
In the Maxwell and Andrade rheological models, $K(\nu)$ is an odd function of the tidal frequency mode $\nu$ with two anti-symmetric peaks typically located close to the crossing point at $\nu=0$ and slopes tending to zero outside the peak frequency modes~\cite{2012ApJ...746..150E}. 
Furthermore, at~higher tidal frequencies, the~function of tidal quality is inversely proportional to the tidal mode, $K(\nu)\propto 1/\nu$. For~an order-of-magnitude estimation, we can ignore the relatively small contributions from the terms proportional to $K(n)$ and $K(2n)$ and omit the integer multiples of $n$ in the arguments of the other terms. The~rate of orbital decay then simplifies to
\begin{equation}
\label{dadt.eq}
\frac{da}{dt} \simeq -\frac{3}{8} n\,\frac{M_N}{M_V} \frac{R_V^5}{a^4}\, \left[8+108\,e^2+573\,e^4\right] K(2\Omega).
\end{equation}

\begin{figure}[H]
\includegraphics[width=220pt]{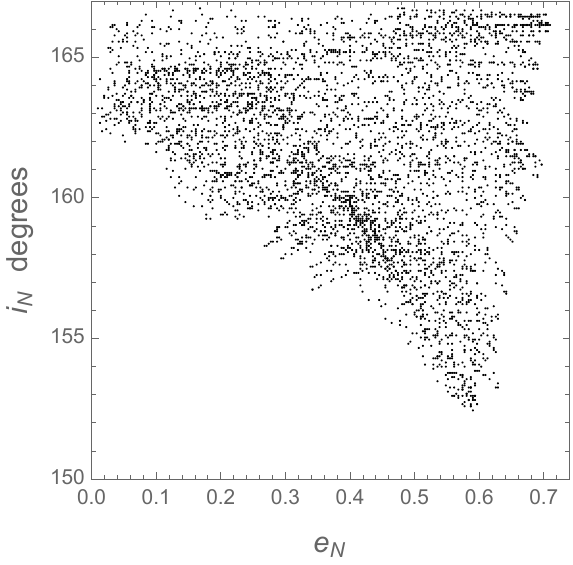}
\caption{Distribution of osculating orbital elements of a chaotically captured moon in a retrograde orbit around Venus in the eccentricity--inclination plane. The~range of inclination variations increases when the osculating eccentricity is~high.} 
\label{ie.fig}
\end{figure}

The orbital elements entering this equation are not fixed values for the chaotically moving Neith but are rapidly varying functions of time. The~osculating orbital elements within the capture stage are computed from the relative coordinates and velocities of Neith in the output of the numerical integration. For~the particular configuration used here, we find a range of 0.01 to 0.71 for the osculating eccentricity, 0.0042 to 0.0064 AU for the semi-major axis, 20 to 36 rad/yr for the mean motion, and~$152$ to $167$ degrees for inclination. The~trajectories of Neith in its retrograde motion around Venus map to complex  hyper-volumes in the space of Kepler elements. Figure~\ref{ie.fig} shows the sampled values of eccentricity and inclination extracted from the simulation output. Despite the strongly chaotic nature of these trajectories, the~osculating elements are loosely interrelated. The~inclination range of Neith narrows down to approximately 4 degrees when the orbit becomes nearly circular, and~its orbital spin shifts toward the planar retrograde configuration ($i_N=180$ degrees). A~cross-section of the osculating elements in the $a_N$--$e_N$ plane reveals that there is a complex pattern of preferred states where the captured moon spends most of the~time.

The tidal quality value in Equation~(\ref{dadt.eq}) is quite uncertain. The~main factor defining this parameter in the Maxwell tidal response model is viscosity, which is likely to be much lower for the freshly formed Venus than for the Venus of today. The~tidal quality is certainly much smaller than unity because~its peak value is limited by the second-degree static Love number. The~coefficient in front of $K(2\Omega)$, which is also a rapidly varying function of time, can be directly computed from our sampled output. It follows a lopsided distribution peaking close to its upper bound, with~a median value of $-1.2\times 10^{-12}$ AU/yr, assuming that the mass of Neith equals the mass of the Moon. The~corresponding e-folding time of tidal orbital shrinkage is approximately 4 Gyr, which is comparable with the age of Venus. The~actual characteristic time can be shorter by an order of magnitude because our approximation is limited to $O(e^4)$, and, as~seen from Equation~(\ref{dadt.eq}), the~omitted higher-degree terms are likely to be dominating. However, the~tidal quality factor is likely to push the estimated rate of tidal decay even lower. The~emerging conclusion is that the tidal dissipation mechanism inside Venus is not sufficient to quickly shrink and stabilize the orbit of the captured~moon.

\section{Close-Range Evolution of Orbit and~Spin}
A giant impact of a Mercury-sized body is the leading theory of Earth--Moon system formation. The~impacting body in this scenario is mostly merged with the early Earth, but~a massive amount of debris from the mantle is swept up to distances above the Roche radius, which serves as construction material for the Moon. Similar ideas can be considered for Neith. Gerstenkorn~\cite{1955ZA.....36..245G} considered a variety of scenarios when the newly formed moon has a retrograde orbital momentum with respect to the angular momentum of the planet. The exchange of angular momentum between the Moon and the Earth results in the orbit flipping and becoming prograde. Here, we consider a different option for Venus and Neith when the moon remains retrograde long enough to reverse the spin of the~planet.

There are two constraints to be taken into consideration. The~tidal orbital decay limits the life span of the retrograde moon. A~retrograde moon is loosing its orbital angular momentum and spirals onto the planet. At~the same time, the~prograde rotation of the planet slows down and may eventually reverse its direction. There is a competition in speed between these two processes~\cite{2023A&A...672A..78M}, and~the end result is either a completely synchronized planet or destruction of the moon when it reaches the Roche radius. From~the conservation of the total angular momentum,
\begin{equation}
    \frac{|\dot\Omega_V|}{|\dot n|}=\frac{M_N}{\xi (M_V+M_N)} \left(\frac{a}{R_V}\right)^2
    \left[ \frac{1}{3}\sqrt{1-e^2}+\frac{|n|}{|\dot n|}\frac{e\,\dot e}{\sqrt{1-e^2}}\right],
\end{equation}
where $\Omega_V$ is Venus' angular velocity of rotation, $\xi$ is the rotational inertia coefficient, and~the upper dots denote time derivatives $d/dt$. In~the initial configuration, $n<0$, $\Omega>0$, and~$\dot\Omega<0$. The~sign of $\dot e$ depends on the initial eccentricity, among~other parameters. It can be negative for small $e$, which effectively slows down the relative rotational deceleration. For~an order of magnitude estimate, we set $e$ to zero. Direct calculations show that the rates of orbital decay and rotational spin-down become equal for $M_N=M_{\rm Moon}$ at $a\approx 8.1\,R_V$. As~long as Neith remains above this critical distance, it rapidly brakes the rotation of Venus trying to synchronize it in the retrograde direction. After~crossing the critical radius, Neith continues to spiral down with ever-increasing speed, while the reversal of Venus' spin slows~down. 

The second constraint comes from the consideration of energy balance. A~spiraling down Neith is a source of energy. The~rotating Venus is also initially a source of energy, as~long as its sidereal rotation is prograde. A~descent of Neith of one Lunar mass \mbox{($M_N=M_{\rm Moon}$)} from $10\,R_V$ to $3\,R_V$ releases $4.7\times 10^{29}$ J of energy, for~example. An~additional $1.4\times 10^{29}$ J is extracted from Venus's rotational energy if it slows down from one revolution per day to zero rotation over the same period of time. The~total amount (of $6.1\times 10^{29}$ J) has been spent on the tidal heating of Venus. Stretched over 100 Kyr, this tidal heating process results in a surface flux of 420 W m$^{-2}$, which is roughly one-fifth of the average solar insolation flux. Thus, the~tidal dissipation of energy extracted from the falling Neith and slowing rotation could provide a significant additional source of heating of Venus' surface. This scenario is plausible if an additional agent could rapidly deliver the captured Neith from the Hill radius to a much closer~range.

\section{Dynamical Friction in the Planetary Debris~Disk}

At the late stages of planetary system formation, the~oligarchic growth of the largest and evenly spaced planetary bodies comes to a halt when most of the smaller bodies have been collisionally fragmented, swept up, or~ejected to the outskirts of the system~\cite{2004ApJ...614..497G}. The~balance is tipped toward coalescence and sweep-up for the inner planets, which have smaller masses and Hill radii. The~cleaning of the inner protoplanetary disk was probably assisted by smaller debris disks of higher mass density revolving around the surviving oligarchs. The~extents of such debris disks are limited to the Hill radius of each planet. For~a disk rotating around Venus in the prograde direction, the~outer radius is about $81 R_{V}$. A~moon captured through the chaotic layer in a retrograde orbit makes often incursions inside this radius because of the stochastic nature of its motion (Figure~\ref{sep.fig}), where it interacts with the particles and small planetesimals. This interaction is also stochastic, but~the net result is the dynamic friction of the orbital motion of the moon. A~retrograde Neith moves against the flow of particles in the prograde disk, roughly doubling the average relative velocity $v_{\rm rel}$. This is a case of shear-dominated friction with the Coulomb \mbox{parameter $\Lambda << 1$ \cite{2007ApJ...661..602F}}. The~friction equation can be written as
\begin{equation}
\frac{dv}{dt}=-\frac{4\sqrt \pi}{\sqrt 2 M_N}\, \sigma \, n \,R^2_{\rm HN} \, v_{\rm rel},
\end{equation}
where $R_{\rm HN}$ is the Hill radius of Neith orbiting around Venus, and~$\sigma$ is the surface density of the disk. The~spiraling-down trajectory of Neith can be obtained by integrating this equation with the specific simulated outputs in terms of the six phase-space parameters. Here, however, we settle on an approximate estimation, assuming that the encounters between the moon and disk particles are elastic, and~that their orbital velocities are Keplerian (with opposite signs). The~induced rate of orbital decay is then
\begin{equation}
\frac{dr}{dt}\simeq-\frac{8\sqrt{2 \pi}}{M_N} \, \sigma \,r\,n\, R^2_{\rm HN},
\end{equation}
where $r$ is the separation between Venus and Neith. This can be further simplified by assuming a uniform distribution of the surface density $\sigma$. After~substituting the formula for $R^2_{\rm HN}$, which is also a function of $r$, one obtains
\begin{equation}
\frac{dr}{dt}\simeq-\frac{8\sqrt{2 \pi\,G\,(M_V+M_N)}}{M_N} \left(\frac{M_N}{M_V}\right)^\frac{2}{3}\sigma \,r^\frac{3}{2}.
\end{equation}

We note that the friction-induced orbital decay is faster for a less massive moon, but~the dependence is relatively weak. This differential equation has an analytical solution, \mbox{which is}
\begin{eqnarray}
    r(t)&=&\left(\frac{1}{\sqrt{r(0)}}+H\,t\right)^{-2}, \nonumber\\
    H&=& \frac{4\sqrt{2 \pi\,G\,(M_V+M_N)}}{M_N} \left(\frac{M_N}{M_V}\right)^\frac{2}{3}\sigma.
\end{eqnarray}

\begin{figure}[H]

\includegraphics[width=220pt]{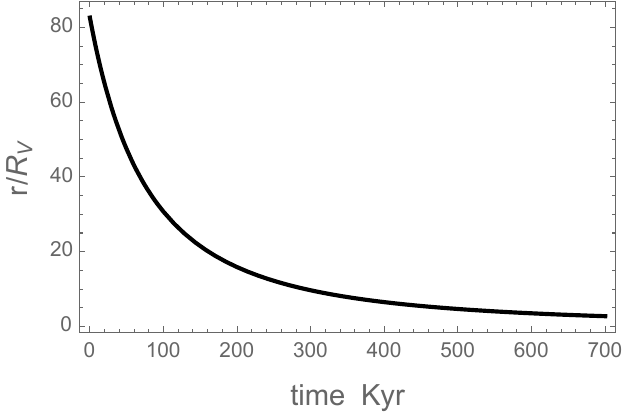}
\caption{{Orbital} 
 decay of a retrograde moon captured by Venus due to shear-dominated dynamical friction in the prograde debris~disk.} 
\label{friction.fig}
\end{figure}

Figure~\ref{friction.fig} displays the computed decrease of the orbital separation between Neith and Venus with the following parameters: $M_N=7.35\times 10^{22}$ kg, $M_V=4.87\times 10^{24}$ kg, $\sigma=0.06$ kg m$^{-2}$. The~initial decline is fast, but~it slows down at close distances to the planet despite the increasing relative velocity because of the shrinking area of gravitational influence. At~the critical distance of $8.1\,R_V$, the~tidal braking takes over, while the angular deceleration of Venus continues to rapidly increase. Thus, dynamical friction in the residual circumplanetary disk can efficiently reduce the angular momentum of an initially distant moon captured into a retrograde orbit. This allows the orbit to stabilize and become less eccentric preventing a secondary stochastic~escape.

\section{Summary}

Venus is the only inner planet in the Solar system involved in a slow retrograde rotation with respect to the orbital motion. It also lacks a permanent moon. Both these features can be explained in the proposed scenario of a sufficiently massive moon (called Neith here) captured through the fractal Hill surface of Venus at the early stages of dynamical evolution immediately following the epoch of oligarchic growth. The~remnants of the protoplanetary disk still included a large number of smaller planetesimals that failed to acquire sufficient mass to become oligarchs and clear out a dynamical niche for long-term survival. Using precision numerical methods, we demonstrate the possibility of capture into a chaotic retrograde orbit, which can last in the conservative regime for $10^5$--$10^6$ years. This happens when the smaller body approaches Venus at the appropriate angle from its heliocentric trajectory and undergoes a dynamical maneuver similar to the gravity assisted deceleration. The~osculating orbital elements of the captured Neith rapidly vary in a wide range. As~a result, the~captured moon makes frequent incursions into the closer area around the planet, occasionally approaching it as close as ten Venus radii. Without~assistance from external dissipative agents, however, Neith is doomed to be ejected again into a heliocentric orbit crossing with Venus and, possibly, with~other~planets.

We show in this paper that the tidal dissipation inside Venus is too slow to reduce the excessive orbital momentum of the temporary moon. The~average separation is still close to the Hill radius for a retrograde particle, and~the rate of tidal orbital decay is approximately proportional to $a^{-11/2}$, where $a$ is the osculating semi-major axis. The~tidal quality modes are also likely to have low values for Venus still rotating in the prograde sense. At~separations above the critical value of $8.1\,R_{V}$, the~rate of rotational braking is higher than the rate of mean motion acceleration, irrespective of the tidal quality. Thus, the~dynamic tidal deformation of Venus can relatively quickly reduce the spin rate of the planet, but~it is not efficient in bringing Neith down to a closer~range.

The situation changes when Neith approaches the critical orbital radius. The~rate of tidal despinning and, eventually, reversal of Venus' rotation accelerates and becomes faster than the tidal component of orbital decay. Equation~(\ref{da.eq}) describing the rate of orbital decay is derived from the massive theoretical work {started} by Darwin and Kaula \citep{1880RSPT..171..713D, 1964RvGSP...2..661K} and further developed by Efroimsky \citep{2012CeMDA.112..283E}, which combines the Fourier expansion of the alternating external gravitational perturbation in harmonics of mean longitude with a frequency-dependent rheological response. The~derived model, however, only captures the polar component of the variable tidal torque, which is further averaged out in time over one orbital period and one period of precession. In~our simulations, Neith is allowed to be captured with an initial orbital inclination deviating from $180^\circ$ relative to the initial Venus' spin axis. The~equatorial components of the tidal torque are strictly periodic, causing the orbital axis of Neith and the spin axis of Venus to librate. The~emerging dissipation of the latitudinal torque makes the orbit and the spin realign toward an equilibrium state in accordance with the preservation of the total angular momentum. The~current obliquity of Venus ($177.3^\circ$) can be the consequence of the initial orbital inclination of the captured~Neith.

The function of removing the excessive orbital momentum of the retrograde Neith and delivering it to the critical separation  within $\sim$$10^5$ years can be performed by a residual debris disk revolving around Venus, which is composed of small pebbles, dust, and~gas. The~stochastic incursions of Neith inside the disk area limited by the prograde Hill radius cause encounters of Neith with the disk particles at a high relative velocity. The~net outcome is shear-dominated dynamical friction, which accelerates the orbital motion by reducing the orbital extent. Our approximate theoretical calculations show that the initial rate of orbital decay by friction is high enough to reduce Neith's orbital momentum to sufficiently low levels. This moves the captured moon closer to Venus within the natural capture cycle and precludes its stochastic escape. At~a distance of a few Venus radii, the~frictional decay drastically slows down because of the shrinking impact parameter, while the tidal decay becomes significant and takes over. This process could be interrupted by a complete synchronization of Venus, which would almost nullify the tidal interaction. However, the~rotational braking also slows down at separations below the critical distance. There is no time left for Neith to synchronize Venus, and~it crashes into the planet when it is slowly rotating in the retrograde~direction.

\vspace{6pt}

\authorcontributions{{Conceptualization, V.V.M. and A.G.; methodology, V.V.M.; software, A.G.; validation, V.V.M.; formal analysis, V.V.M. and A.G.; investigation, V.V.M. and A.G.; resources, A.G.; writing---original draft preparation, V.V.M.; writing---review and editing, V.V.M. and A.G.; visualization, V.V.M.; supervision, V.V.M. All authors have read and agreed to the published version of the~manuscript.} 
}

\funding{{This research received no external~funding.} 
}

\dataavailability{The simulation inputs and results discussed in this paper are available upon reasonable request to the corresponding {author.} }

\acknowledgments{{The authors thank Michael Efroimsky for inspiring the discussions on the topic of this paper and sharing his insight into the theory of~tides.}} 

\conflictsofinterest{{The authors declare no conflict of interest.} 
}

\begin{adjustwidth}{-\extralength}{0cm}

\reftitle{{References} 
}


\PublishersNote{}
\end{adjustwidth}

\label{lastpage}
\end{document}